\documentclass[namedreferences]{SolarPhysics}
\usepackage[optionalrh]{spr-sola-addons} 
\usepackage{graphicx}        
\usepackage{color}           
\usepackage{url}             

\begin{document}

\begin{article}

\begin{opening}

\title{Influence of Low-Degree High-Order {\bf \emph{p}}-Mode Splittings on the Solar
      Rotation Profile} 

\author{R.A.~\surname{Garc\'\i a}$^{1}$\sep
        S.~\surname{Mathur}$^{1}$\sep
        J.~\surname{Ballot}$^{2}$\sep
        A.~\surname{Eff-Darwich}$^{3,4}$\sep
        S.J.~\surname{Jim\'enez-Reyes}$^4$\sep
        S.G.~\surname{Korzennik}$^{5}$
       }
\runningauthor{R.A~Garc\'\i a \emph{et al.}}
\runningtitle{Influence of Low-$\ell$ High-$n$ $p$ Modes Splittings 
              on the Rotation Profile}

\institute{
 $^{1}$ Laboratoire AIM, CEA/DSM-CNRS - U. Paris Diderot - IRFU/SAp, 
91191 Gif-sur-Yvette Cedex, France  \url{rgarcia@cea.fr}, 
\url{smathur@cea.fr}\\ 
 $^{2}$Max-Planck-Institut f\"ur Astrophysik, Karl-Schwarzschild-Strasse 1, 
85748 Garching, Germany  \url{jballot@mpa-Garching.mpg.de}\\
 $^{3}$ Departamento de Edafolog\'\i a y Geolog\'\i a, 
Universidad de La Laguna, Tenerife, Spain \url{adarwich@ull.es} \\
 $^{4}$ Instituto de Astrof\'\i sica de Canarias, 38205, 
La Laguna, Tenerife, Spain \url{adarwich@iac.es}, \url{sjimenez@iac.es} \\
 $^{5}$Harvard-Smithsonian Center for Astrophysics, 60 Garden Street, 
Cambridge, MA 02138, USA \url{skorzennik@cfa.harvard.edu}
             }

\date{Received 10 October 2007; accepted 31 December 2007}

\begin{abstract}

The solar rotation profile is well constrained down to about 0.25 $R_{\odot}$
thanks to the study of acoustic modes. Since the radius of the inner turning
point of a resonant acoustic mode is inversely proportional to the ratio of
its frequency to its degree, only the low-degree $p$ modes reach the core. The
higher the order of these modes, the deeper they penetrate into the Sun and
thus they carry more diagnostic information on the inner
regions. Unfortunately, the estimates of frequency splittings at high frequency
from Sun-as-a-star measurements have higher observational errors due to mode
blending, resulting in weaker constraints on the rotation profile in the inner
core. Therefore inversions for the solar internal rotation use only modes below
2.4 mHz for $\ell \le 3$. In the work presented here, we used an 11.5 year-long time series to compute the rotational frequency splittings for modes
$\ell \le 3$ using velocities measured with the GOLF instrument. We carried
out a theoretical study of the influence of the low-degree modes in the region
2 to 3.5 mHz on the inferred rotation profile as a function of their error
bars.

\end{abstract}
\end{opening}

\section{Introduction}
     \label{S-Introduction} 
 Our knowledge of the solar rotation profile has been derived from the study
of the resonant acoustic modes, which are trapped in the solar interior. Since
the solar rotation lifts the azimuthal degeneracy of these resonant modes,
their eigenfrequencies ($\nu_{n \ell m}$) are split into their $m$-components;
where $\ell$ is the angular degree, $n$ the radial order, and $m$ the
azimuthal order. This separation $\Delta \nu _{n \ell m}$ --- usually called
rotational splitting (or just splitting) --- depends on the rotation rate in
the region sampled by the mode. Using inversion techniques, the rotation rate
at different locations inside the Sun can be inferred from a suitable linear
combination of the measured rotational splittings
 
Today, the rotation rate inside the Sun is rather well known above 0.4
$R_\odot$
\cite{ThoTOO1996,1998ApJ...505..390S,HowJCD2000,2000ApJ...541..442A}. The
convective zone is characterized by a differential rotation extending from the
surface down to the tachocline, located around 0.7 $R_\odot$. Below the
tachocline, inside the radiative region, the Sun appears to rotate as a rigid
body with a nearly constant rate of $\approx$ 433~nHz down to the solar core,
{\em i.e.} $\approx$ 0.25 $R_\odot$. The rotation rate inside the core derived
from $p$ modes is still uncertain
\cite{1994ApJ...435..874J,ElsHow1995,ChaEls2001,ChaSek2004}. Recent
measurements of the asymptotic properties of the dipole $g$ modes and their
comparison with solar models favors a faster rotation rate inside the solar
core \cite{2007Sci...316.1591G}.

For $p$ modes of a given degree $\ell$, the radius at the inner turning point
is a decreasing function of frequency, given by:
\[
r_t = c_t\, L/( 2 \pi \nu_{n \ell})
\]
where $L=\ell+1/2$, $\nu_{n \ell}$ is the frequency 
of the mode, and $c_t = c(r_t)$ the sound-speed at the radius $r_t$ (see for
example, \opencite{1994A&A...290..845L}). Thus the modes with increasing
frequencies -- higher radial order $n$ -- penetrate deeper inside the
Sun. Unfortunately, when fitting Sun-as-a-star observations the uncertainties
on the rotational splittings that are the most sensitive to low-degree $p$ modes
are very large. Indeed, as the modes lifetimes decrease with frequency their
line widths increase. Therefore, for frequencies above $\approx2.3$ mHz, there is
a substantial blending between the visible $m$-components of the $p$ modes. This
blending makes it difficult to extract precisely the rotational splitting. At
higher frequencies ($\approx3.9$ mHz) even the successive pairs of $\ell=0, 2$
and $\ell=1, 3$ modes blend together. As a result, with today's fitting
methods, it is still not possible to obtain values of the rotational
splittings with an accuracy good enough to be useful in any rotation
inversion.



  By contrast, at low frequency -- below $n=16$, or about 2.4 mHz -- the
lifetime of the modes increases and thus their line width is very small. This
allows us to measure their rotational splittings with very high
precision. However, these modes have inner turning points at shallower depths
than the high-frequency modes (above 0.08 and 0.12 $R_\odot$ for the $\ell=1$
and 2 modes respectively). Therefore, even though these modes do not carry any
information below $\approx 0.1$ $R_\odot$, they help improve our knowledge of the
inner rotation rate because their inclusion contributes to an increase of the
precision of the inversions since they have smaller error bars
\cite{2002ApJ...573..857E,effKor07}.

  For all of these reasons, rotation inversion methodologies usually limit the
input data set of low-degree $p$ modes to low frequency modes.  For example,
\inlinecite{CouGar2003} used a limited number of splittings, corresponding to
modes with $\ell \le 3$ and frequencies below 2.4 mHz ($n=15$) resulting from
fitting GOLF\footnote{Global Oscillations at Low Frequency \cite{GabGre1995}}
and MDI\footnote{Michelson Doppler Imager \cite{1995SoPh..162..129S}} 2,243
day-long velocity time series, to infer the solar rotation profile.  He
concluded that the uncertainties in the rotation rate below $0.3 R_\odot$ were
still quite large. Therefore, to obtain a better and more reliable rotation
profile in the inner core, we need, on one hand, to include low-frequency
acoustic, mixed, and gravity modes in the inversions (see discussions in
\opencite{ProBer2000} and \opencite{MatEff2007Inv}). On the other hand, we
also need to measure more accurately low-degree high-order $p$ modes to push
further the frequency limit of the modes used in the inversions.

  We present, in Section 2, rotational splittings of $\ell \le 3$ modes
computed using 11.5 years of GOLF data, paying special attention that no bias
is introduced by the solar activity or by the length of the fitting
window. Thanks to the length of these new time series, the splitting error
bars are improved by $\sqrt T$ when compared to previous and shorter analysis
using the same instrument. In Section 3, we study the sensitivity of the
rotation rate below 0.25 $R_\odot$ to the available $p$ modes and their error
bars. We then discuss, in Section 4, the influence of these modes on the
inversions, based on the resulting resolution kernels and we finish by
discussing the inverted rotation profile using the newly computed GOLF
splittings.


\section{Data Analysis}

 We analyzed a $4\,182$ day-long time series of GOLF observations that spans the
11 April 1996 to 22 September 2007 epoch ($\approx$ 11.5 years). These
observations were calibrated following the methods described in
\inlinecite{GarSTC2005}. The duty cycle of that time series is 94.5$\%$. The
GOLF instrument is a resonant scattering spectrophotometer onboard the
SOHO\footnote{SOlar and Heliospheric Observatory \cite{DomFle1995}}
spacecraft. A standard Fast Fourier Transform algorithm was used to
compute the power spectral density. This periodogram estimator was used
because our primary interest is the study of modes in the medium frequency
range. We have therefore decided to privilege the simplest estimator, avoiding
the use of multi-tapers or zero-padding estimators that are better suited to
look for weak and narrow peaks at low frequency.
 
  To obtain the mode parameters, each pair of modes ($\ell=0, 2$ and
$\ell=1,3$) was fitted to a set of asymmetrical Lorentzian, as defined in
\inlinecite{1998ApJ...505L..51N}, using a maximum-likelihood method since the
power-spectrum estimator follows a $\chi^2$ distribution with two degrees of
freedom.  For additional details on this method see, for example,
\inlinecite{1994A&A...289..649T}.  For each mode, we fitted the following set of
parameters: the amplitude, the line width, the central frequency, the peak
asymmetry, the rotational splitting (if $\ell >$ 0), and the 
background noise.

It has been shown that the amplitude ratios of the $m$ components inside a
multiplet could produce a bias in the splitting determination if they are not
correctly set \cite{ChaApp2006}. We have therefore fixed them to the averaged
values obtained directly from the GOLF data set itself (for details see
\opencite{1999PhDT.........8H}). Finally, to reduce the parameter space, we
used the same asymmetry and line width for both modes in any given fitting
window. Indeed \inlinecite{ChaApp2006} demonstrated that using the same width
for the modes fitted simultaneously in a given fitting window reduces the bias
in the splitting determination at high frequencies. Finally, the uncertainties
on the fitted parameters were derived from the square root of the diagonal
elements of the inverted Hessian matrix.

\begin{figure}[h]   
  \centerline{\includegraphics[width=1.10\textwidth,clip=]{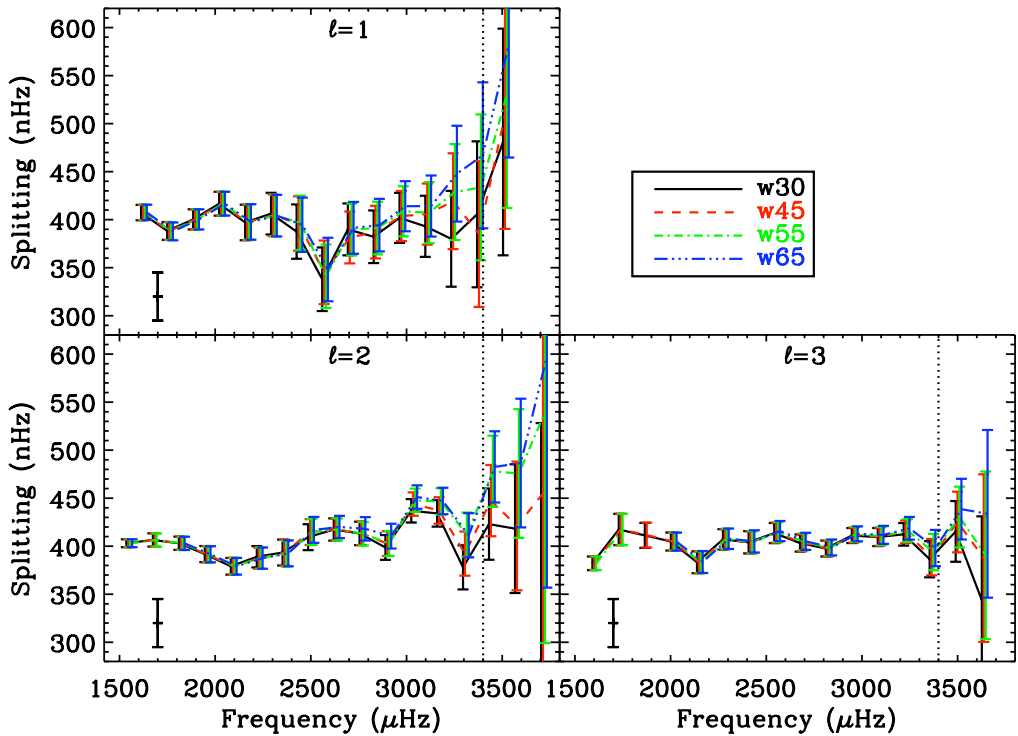} }
  \caption{Synodic splittings for the $\ell= 1, 2$, and 3 modes fitted using
   increasing fitting window widths: from 30 to 65 $\mu$Hz. The vertical
   dotted lines mark the upper limit of the modes whose splittings are used in
   this work. The error bar in the bottom-left corner of each plot corresponds
   to $\pm$ 20 nHz and is plotted as reference.}
   \label{splitw}
\end{figure}


It has been shown, using artificial simulations, that the determination of the
rotational splitting could be biased by the width of the fitting window
\cite{ChaApp2006}. This is caused by the leakage of the neighborin$g$ modes. For
example the splittings of the $\ell=2$ modes could be noticeably modified
(increased) at frequencies above 2.5 mHz due to the leakage of the $\ell=4$
modes. We have  studied this effect using the GOLF data by selecting 4
different windows for the fit: 30, 45, 55, and 65 $\mu$Hz. The results are
plotted in Figure~\ref{splitw}. As expected, the values fitted for the $\ell=2$
splittings increase at high frequencies when wider fitting windows are
used. This behavior is consistent with the results obtained by the solarFLAG
group (see for comparison Figure 9 of \opencite{ChaApp2006}). However,
the smallest window (30 $\mu$Hz), also biases the fittings at high frequency
because fitting such narrow spectral range prevents the fitting code from
properly constraining the background level. Therefore, we have adopted the 45
$\mu$Hz window as the best compromise between these two effects. In the case
of the $\ell =1$ modes the splittings computed using the smaller windows are
roughly constant up to 3.4 mHz and then increase.  For higher frequencies the
splittings increase whatever window width is used and it is not possible to
distinguish between a bias induced by the blending of the two adjacent
$m$-components or a real increase. Therefore we have limited our study to the
splittings of modes below 3.4 mHz (this limit is indicated with a vertical
dotted line in Figure~\ref{splitw}) where all of the splittings are roughly
constant within the error bars. Another approach to limit the leakage of
the neighbor peaks could be to fit the whole spectrum at once with a simple model of the background noise \cite{1998ESASP.418..323R}.

It is also important to notice that some splittings (for example the $\ell=1$
$n=17$ mode) are slightly out of the general trend of the rest of the
splittings (less than 2 $\sigma$) as a consequence of the stochastic nature of
the excitation. This deviation is in quantitative agreement with the results
obtained from artificial simulations (see \opencite{ChaApp2006}).

 We have also verified that there is no noticeable influence of the solar
activity on the extracted splittings. To do this, we followed an approach
similar to the one described in \inlinecite{2001A&A...379..622J}: we divided
the time series in 12 independent 350 day-long segments\footnote{one 350
day-long segment was not fitted since its duty cycle is very small because of
the loss of contact with the SOHO spacecraft during that epoch} and fitted
them simultaneously. A linear dependence on a solar activity index was added
to each parameter defining the spectral profiles, with two exceptions: first,
we fixed the asymmetries to the values found by analyzing the full time
series; second, we included a quadratic dependence on activity for the
frequencies of the modes. 
We used the integrated 10.7 cm radio flux measurements from the National
Geophysical Data Center\footnote{\url{http://www.ngdc.noaa.gov/ngdc.html}}, averaged
over the duration of each sub-series as indicator of solar activity.  The 10.7
cm radio flux has been shown to be the indicator that best correlates with the
measured changes resulting from solar activity.

\begin{figure}
  \centerline{\includegraphics[width=0.52\textwidth,clip=]{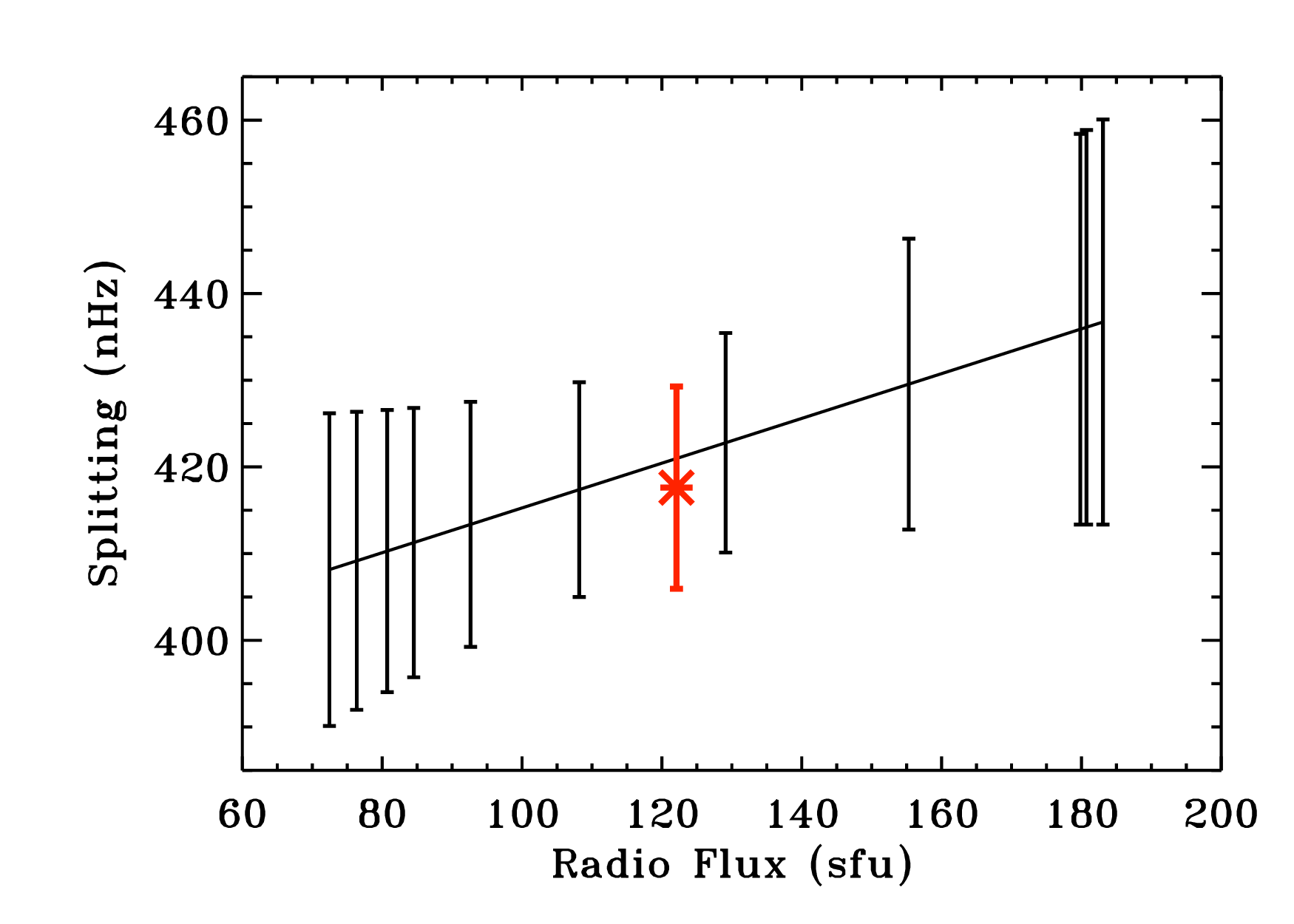}\includegraphics[width=0.52\textwidth,clip=]{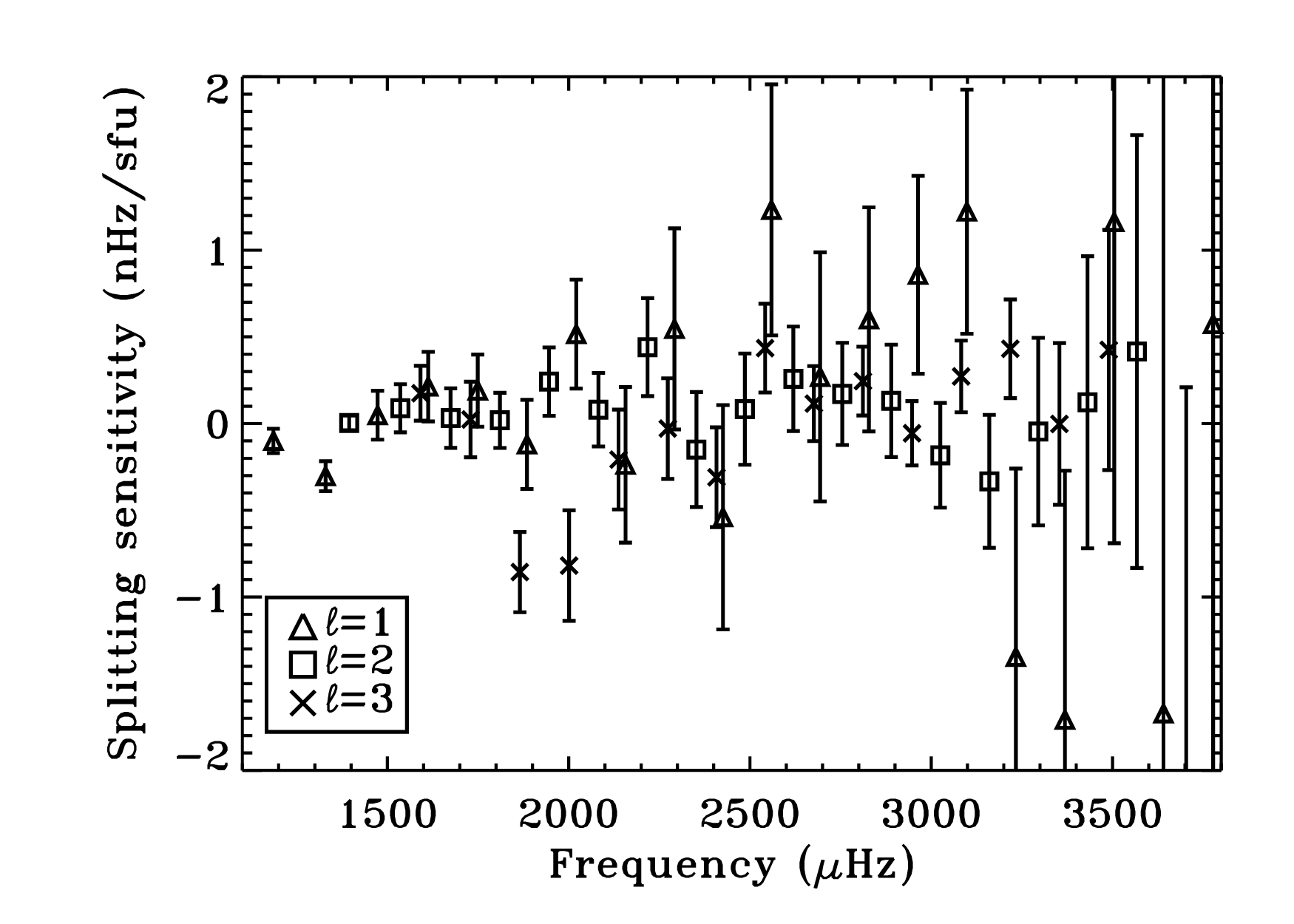}
}
  \vspace{-0.32\textwidth}   
  \centerline{ \small    
  \hspace{0.075\textwidth}\color{black}{(a)}
  \hspace{0.48\textwidth}\color{black}{(b)}
  \hfill}
  \vspace{0.28\textwidth}    

  \caption{(a) Fitted splittings for the mode $(\ell=2,n=17)$ as a function of
  the solar activity level measured by the solar radio flux at 10.7\,cm ($1\,\mathrm{sfu} = 10^{-22}\mathrm{W m^{-2} Hz^{-1}}$). The
  asterisk with thick error bar indicates the result of fitting the full time
  series. (b) Sensitivity of the splittings to the activity for all of the
  fitted modes ($\ell=1,2$ and 3).}
  \label{activity}
\end{figure}

Figure~\ref{activity}a illustrates the variation of the splitting with solar
activity for one mode ($\ell=2$, $n=17$), while Figure~\ref{activity}b shows
the sensitivity of the splittings to the radio flux for all of the fitted
modes ({\em i.e.} the slope in Figure~\ref{activity}a).  This plot shows
that there is no clear systematic dependence like the one seen for the
frequency
\cite{2002SoPh..209..247J,2002A&A...394..285G,2004SOHO...14..436G}. Moreover
variations during the solar cycle are generally marginally significant.
The splittings obtained by fitting the whole time series are fully consistent
with those obtained when using this activity-dependent method and correspond
to the Sun at its mean activity level, as expected.
  The resulting rotational splittings for modes $\ell \le 3$ are listed in
Table~\ref{tabsplit}.
 
  Before inverting these rotational frequency splittings we completed the set
of splittings of low-degree low-order $p$ modes down to 1 mHz with those
extracted from the analysis of combined GOLF and MDI time series
\cite{GarCor2004}. Then we added high-degree modes using splittings of modes
between $\ell=4 $ and $\ell=25$ from the analysis of the $2\,088$ day-long time
series of MDI observations fitted by \inlinecite{Kor2005}.

\begin{table}
 \caption{Central frequencies ($\nu_o$), synodic sectoral splittings ($\Delta
\nu _{n \ell m}$), and their respective 1 $\sigma$ error bars ($\sigma_o$
and $\sigma_{n\ell m}$) computed using $4\,182$ days of GOLF velocity time series.}
\label{tabsplit}
\begin{tabular}{crcrrrr}     
  \hline                   
$\ell$ & n & $\nu_o$ & $\sigma_o$ & $\Delta \nu _{n \ell m}$ & $\sigma_{n \ell m}$  \\
         &    &($\mu$Hz)& (nHz) & (nHz) & (nHz) \\
  \hline
1   & 7  & 1185.5893& 4.0  &  402.2   & 3.2 \\
	& 8  & 1329.6368 & 3.3 &  405.2   & 2.9\\
	& 9  & 1472.8475 & 5.0 &  400.0   & 4.2\\
	& 10  & 1612.7268 & 9.9 & 407.6   & 8.0\\
	& 11  & 1749.2887 & 10.5& 388.1   & 9.2\\
	& 12  & 1885.0853 & 12.2& 400.4   & 10.6\\
	& 13  & 2020.8228 & 14.1& 416.5   & 12.5\\
	& 14  & 2156.8158 & 18.5& 397.3   & 18.4\\
	& 15  & 2292.0340 & 19.0& 404.7   & 21.6\\
	& 16  & 2425.6381 & 20.0& 396.0   & 26.3\\
	& 17  & 2559.2441 & 19.3& 345.2   & 30.7\\
	& 18  & 2693.4385 & 18.5& 381.5   & 27.0\\
	& 19  & 2828.2585 & 18.0& 387.1   & 26.0\\
	& 20  & 2963.4225 & 17.7& 403.8   & 24.9\\
	& 21  & 3098.2937 & 19.0& 405.8   & 29.3\\
	& 22  & 3233.2855 & 22.2& 419.2   & 41.6\\
	& 23  & 3368.6923 & 26.9& 385.1   & 75.6\\
\hline	
2   	&   8  & 1394.6851& 14.2& 401.8   &5.9\\
	&9 & 1535.8642 & 7.0& 403.1    &4.1\\
	&10 &1674.5434 & 12.3& 406.5    &6.9\\
	&11 & 1810.3293& 13.3& 403.0    &6.9\\
	&12 & 1945.8173& 15.9& 391.6    &8.4\\
	&13 & 2082.1175 & 18.5& 379.2    &8.6\\
	&14 & 2217.6968& 22.2& 387.4    &11.3\\
	&15 & 2352.2626& 23.7& 393.4    &13.5\\\
	&16 & 2485.9310& 22.8& 414.7    &12.9\\
	&17 &2619.7165 & 20.6& 417.6    &11.7\\
	&18 & 2754.5761& 20.4& 413.5    &12.2\\
	&19 & 2889.6857& 20.5& 402.7    &12.7\\
	&20 & 3024.8519& 20.3 & 443.9    &11.7\\
	&21 & 3159.9752& 22.3 & 437.2    &13.6\\
	&22 & 3295.2381& 28.1& 392.3    & 21.4\\
	&23 & 3430.9280& 38.6 & 447.5    &33.4\\
\hline       
3	&9 &1591.4891 & 19.5& 382.3  &  7.1\\
	&10 & 1729.1460& 47.7& 417.5    &16.3\\
	&11 & 1865.3041& 36.9& 411.6    &12.9\\
	&12 & 2001.2566& 28.6& 404.9    &9.4\\
	&13 & 2137.7634& 38.2& 383.3    &11.4\\
	&14 & 2273.4932& 32.7& 407.4    &10.6\\
	&15 & 2407.6669& 36.2& 403.8    &11.5\\
	&16 & 2541.7150& 32.7& 414.1    &10.4\\
	&17 & 2676.2540& 27.1& 403.7    &8.9\\
	&18 & 2811.4796& 24.1& 397.5    &8.3\\
	&19 & 2947.1109& 23.1& 411.6    &7.8\\
	&20 & 3082.4893& 27.2& 409.9    &9.3\\
	&21 & 3217.9177& 31.3& 415.0    &11.4\\
	&22 & 3353.6765& 46.1& 388.6    &19.0\\
	&23 & 3489.6852 & 66.4& 425.2    &30.0\\	
  \hline
\end{tabular}
\end{table}

\section{Sensitivity of the Splittings to the Rotation Rate below 0.25
  $R_\odot$} 

\begin{figure}[h]  
  \centerline{\includegraphics[width=0.8\textwidth,clip=]{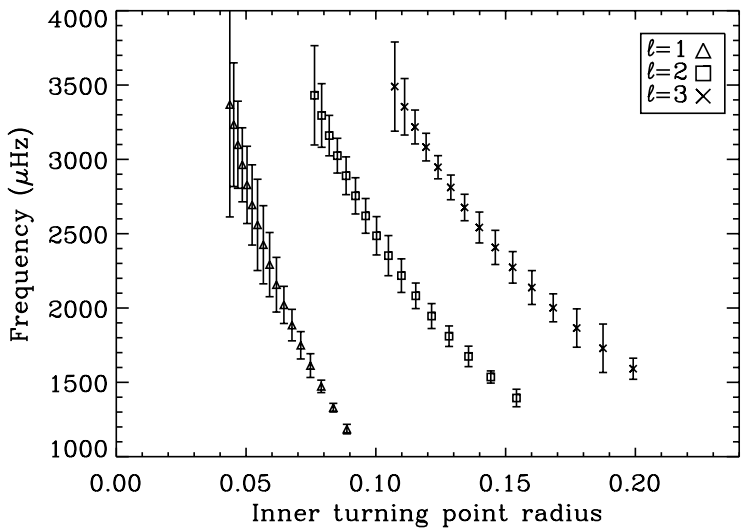} }
  \caption{Fitted modes listed in Table~\ref{tabsplit} as a function of the
  inner turning point radii. The error bars are the splitting error bars in
  nHz magnified a factor $10^4$.}
  \label{turning}
\end{figure}

We have shown in the introduction that the acoustic modes of higher radial
order penetrate deeper in the solar interior and that they are potentially of
great interest to better constrain the rotation inside the solar core.  This
is illustrated in Figure~\ref{turning} where the modes listed in
Table~\ref{tabsplit} are plotted as a function of the radius of their inner
turning point. 

The vertical error bars correspond to the 1 $\sigma$ error bars
of the splittings. These uncertainties are smaller for the $\ell=2$ and 3 than
for the $\ell=1$ modes because the visible $m$ components of the latter are
closer in frequency and therefore blend together at lower frequencies than for
higher-degree modes.

\begin{figure}[h]   
   \centerline{\hspace*{0.015\textwidth}
               \includegraphics[width=0.55\textwidth,clip=]{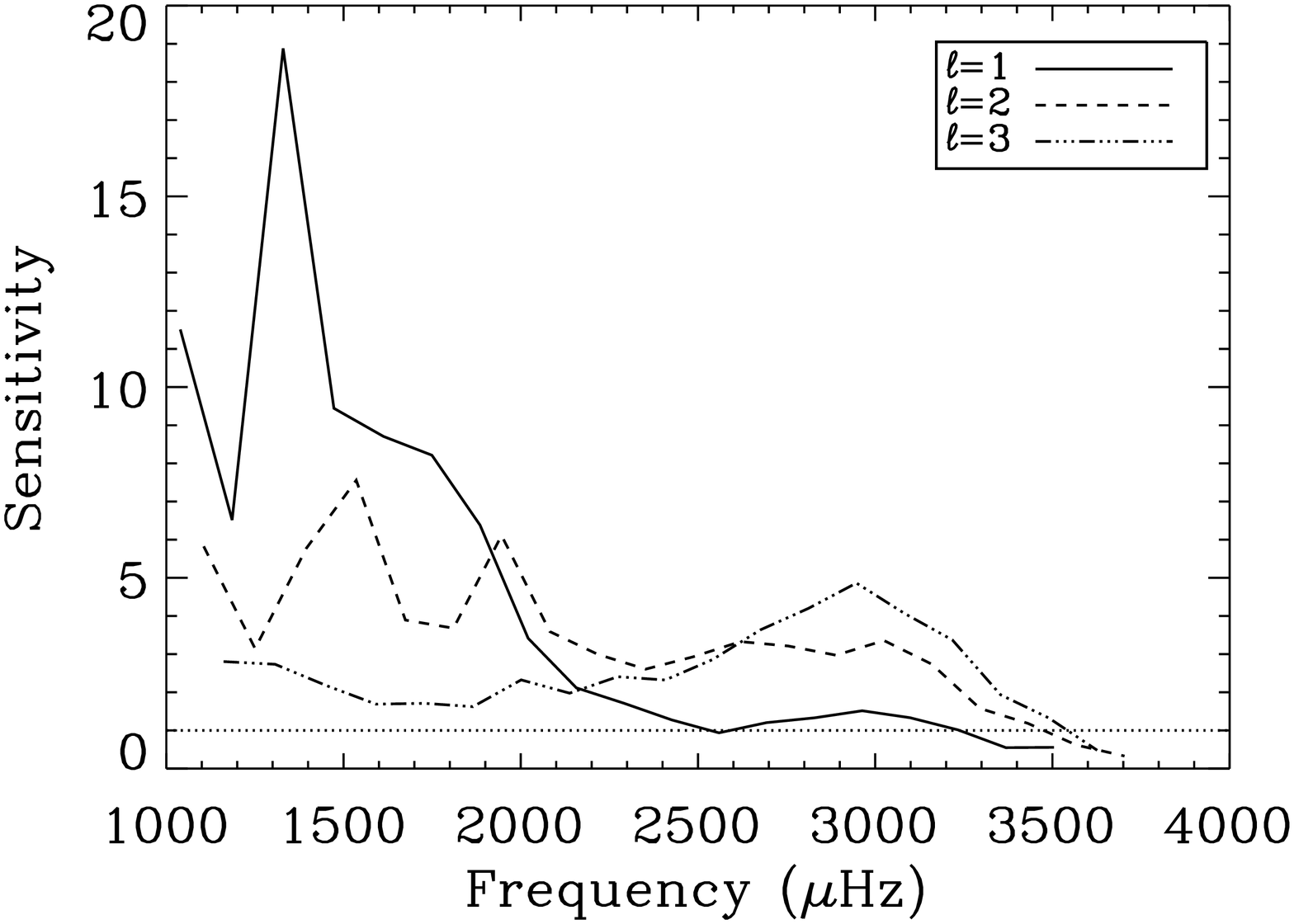}
               \hspace*{-0.05\textwidth}
               \includegraphics[width=0.55\textwidth,clip=]{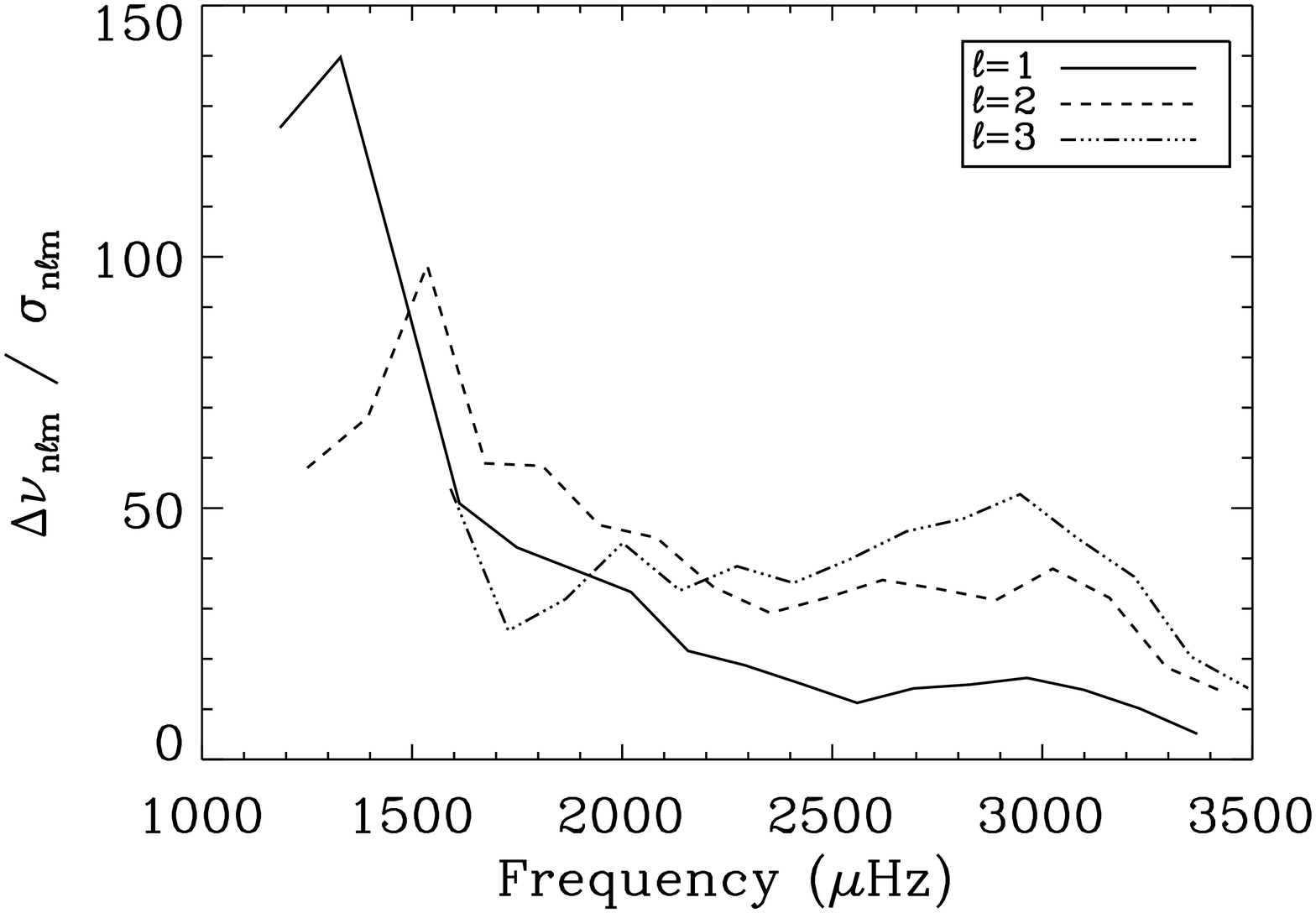} 
              }
    \vspace{-0.35\textwidth}   
     \centerline{ \small    
      \hspace{0.12 \textwidth}  \color{black}{(a)}
      \hspace{0.485\textwidth}  \color{black}{(b)}
         \hfill}
                \vspace{0.31\textwidth}    

 \caption{(a) Sensitivity of the splittings to the rotation rate below 0.25
  $R_\odot$ as defined in the text. The horizontal dotted line is the lower
  limit below which the splittings would not carry any useful information on
  the solar core rotation. (b) Splittings normalized by their error bars.}
  \label{sensib}
\end{figure}

The amount of information on the rotation rate inside the core ($r \le 0.25\,
R_\odot$) present in these modes depends on the precision of the measured
splittings.  The sensitivity of the splittings to the rotation rate below 0.25
$R_\odot$ is plotted in Figure~\ref{sensib}a. This sensitivity is defined as
the ratio of the contribution of the rotation rate below 0.25 $R_\odot$
to the splitting,( $\Delta \nu_{n \ell m}$) and the uncertainty ($\sigma_{n
\ell m}$) in its determination (see equation~\ref{sensi} for sectoral modes, $\ell=m$).

\begin{equation}
Sensitivity = \bigg( \frac{\int_0^{0.25R_\odot} K_{n,\ell}(r) \Omega(r)}{\int_0^{R_\odot} K_{n,\ell}(r) \Omega(r)} \Delta \nu_{n, \ell} \bigg) / \sigma_{n,\ell}
\label{sensi}
\end{equation}

For a given mode, if the sensitivity function
is greater than one, this particular mode provides useful information on the core
rotation. Up to $\approx 3.4$ mHz, all of the modes $\ell \le 3$ are potentially
interesting for the inversions. The situation has evolved drastically since
2003 when the previous analysis of $2\,243$ day-long GOLF and MDI time series
showed that below 0.2 $R_\odot$ only those modes below 2.2 mHz had a
sensitivity greater than one (see left panel of Figure 1 in
\opencite{CouGar2003}). In fact, when we compute the sensitivity function
below the latter radius using the new computed GOLF splittings we obtain the
same dependence with the frequency below 0.25 $R_\odot$ but rescaled to
lower sensitivity. For the $\ell=2$ and 3 modes the sensitivity reduction is
about a factor of two. In the case of the $\ell=1$ modes the factor is smaller and
the sensitivity curve crosses unity at 3.1 instead of 3.3 mHz.

Figure~\ref{sensib}{b} shows the resulting splittings normalized by their
error bars. The precision clearly increases towards the low-frequency range
for the $\ell=1$ modes while for the $\ell=2$ and 3 modes the curves are
flatter indicating that the error bars are nearly constant all along the
analyzed frequency range.

\section{Discussion}

To study the effect of the $\ell \le 3$ modes at medium-range frequencies on
the inferred rotation rate we carry out some 2D inversions.

The rotational splittings ($\Delta \nu_{n \ell m}$) are the integral of
the product of a sensitivity function -- or a kernel -- $K_{n \ell
m}(r,\theta)$, a known function based on a solar model, with the rotation
rate ($\Omega(r,\theta)$) over the radius ($r$) and the co-latitude ($\theta$)
\cite{bi46}:
\begin{equation} 
\Delta \nu _{n \ell m} = \frac{1}{2\pi}\int_0^R \int_0^{\pi}
K_{n \ell m}(r,\theta)\Omega(r,\theta){\rm d}r{\rm d}\theta + \epsilon_{n \ell m}
\label{eq:equation4} 
\end{equation}
where $\epsilon_{n\ell m}$ is the effective error on the measured value
$\Delta\nu_{n\ell m}$. We assume that the errors ($\epsilon_{n\ell m}$) follow
a normal distribution with a standard deviation $\sigma_{n\ell m}$, which is
estimated by the fitting procedure (see Section~2), as listed in Table 1.  This
set of equations defines a classical inverse problem for the solar
rotation. The inversion of this set of $M$ integral equations -- one for each
measured $\Delta \nu _{n \ell m}$ -- allows us to infer the rotation rate
profile as a function of radius and latitude from a set of observed
splittings.

The inversion method used here is based on the regularized least-squares
methodology (RLS) following the prescription described in
\inlinecite{Eff-Darwich1997}. In this implementation the regularization
function is weighted differently for each model grid point (see also
\opencite{effKor07}). In summary, Equation~(\ref{eq:equation4}) is
transformed into a matrix relation:
\begin{equation}
  D = A x + \epsilon \label{eq:equation5}
\end{equation}
where $D$ is the data vector, with elements $\Delta \nu _{n \ell m}$ and
dimension $M$, $x$ is the solution vector to be determined at $N$ model grid
points, $A$ is the matrix with the kernels, of dimension $M \times N$, and
$\epsilon$ is the vector containing the error bars.

The RLS solution for the vector $x$ is given by:
\begin{equation}
  x_{\rm est} = (A^TA + \gamma H)^{-1}A^TD  \label{eq:equation6}
\end{equation}
where $\gamma$ is a scalar introduced to give a suitable weight to the
constraint matrix $H$ on the solution. Replacing $D$ from
Equation~(\ref{eq:equation5}) we obtain

\begin{equation}
  x_{\rm est} = (A^TA + \gamma H)^{-1}A^TAx \equiv Rx  
\label{eq:equation7}
\end{equation}
hence
\begin{equation}
R = (A^TA + \gamma H)^{-1}A^TA \label{eq:equation8} \;.
\end{equation}

The matrix $R$ is referred to as the resolution or sensitivity
matrix. Ideally, $R$ would be the identity matrix, which corresponds to
perfect resolution. However, if we try to find an inverse with a resolution
matrix $R$ close to the identity matrix, the solution is generally dominated
by noise magnification. The individual columns of $R$ display how
anomalies in the corresponding model are imaged by the combined effect of
measurement and inversion. In this sense, each element $R_{ij}$ reveals how
much of the anomaly in the $j^{th}$ inversion model grid point is transferred
into the $i^{th}$ grid point. Consequently, the diagonal elements $R_{ii}$
indicate how much of the information is saved in the model estimate and may be
interpreted as the resolvability or sensitivity of $x_{i}$. We defined the
sensitivity $\lambda_{i}$ of the grid point $x_{i}$ to the inversion process
as follow:

\begin{equation}
\lambda_i = \frac{R_{ii}}{\sum_{j=1}^{N}R_{ij}} \label{eq:equation10}
\end{equation}

With this definition, a lower value of $\lambda_i$ means a lower sensitivity
of $x_{i}$ to the inversion of the solar rotation. We define a smoothing
vector $W$ with elements $w_i=\lambda_i^{-1}$ that is introduced in
Equation~(\ref{eq:equation6}) to complement the smoothing parameter $\gamma$,
namely

\begin{equation}
  x_{\rm est} = (A^TA + \gamma W H)^{-1}A^TD  \label{eq:equation11} \; .
\end{equation}

Such substitution allows us to apply different regularizations to different
model grid points $x_{i}$, whose sensitivities depend on the data set that is
used in the inversions. A set of results can be calculated for different
values of $\gamma$, the optimal solution being the one with the best trade-off
between error propagation and the quadratic difference $\chi^2=|Ax_{\rm
est}-D|^2$ as discussed in \inlinecite{Eff-Darwich1997}.

\begin{table}[h]
\caption{Description of the artificial data sets used to study the sensitivity
of low-degree high-order $p$ modes.}
\begin{tabular}{ccccc}
\hline
 Data set &  $\ell=1$  (mHz) & $\ell=2, 3$  (mHz) &  $\ell > 3$ (mHz)\\\hline
 1 & $1\le\nu\le2.3  $& $1\le\nu\le2.3  $ & $1\le\nu\le3.9$  \\
 2 & $1\le\nu\le2.5  $&  $1\le\nu\le3.4$ & $1\le\nu\le3.9$  \\
 3 & $1\le\nu\le3.4  $&  $1\le\nu\le3.4$ & $1\le\nu\le3.9$  \\\hline \\
      \end{tabular}
    \label{tabsets}
\end{table}
We have performed a theoretical study to determine the effect of adding the
low-degree high-order $p$ modes in the inversions. To do so, we have computed,
using Equation~(\ref{eq:equation4}), the splittings 
corresponding to an artificial rotation profile ($\Omega(r,\theta)$). This
artificial profile has a differential rotation in the convection zone (that
mimics the real one), a rigid rotation from 0.7 down to 0.2 $R_\odot$ equal to
$\Omega_{rz}=433$ nHz and a step-like profile in the core having a rate of 350
nHz in the 0.1--0.2 $R_\odot$ region and a rate three times larger than the rest of
the radiative zone below 0.1 $R_\odot$. Although this profile with steep
changes  and a small drop followed by an increase is unlikely to be
realistic, it enables us to characterize the quality of the inversions as the
inversion code has difficulty reproducing these steep gradients.

\begin{figure}[ht]  
   \centerline{\includegraphics[width=0.75\textwidth,clip=]{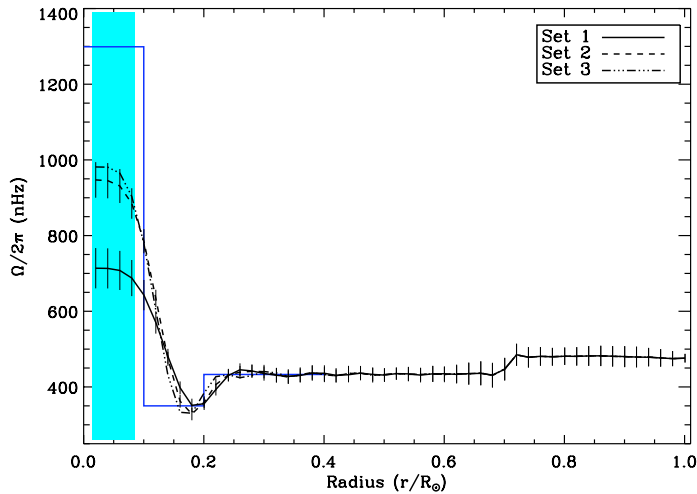} }
 \caption{Equatorial rotation rate obtained from {\em ideal} inversions (all
 the modes with the same error bars) of an artificial step profile (continuous line) using the three data sets described in
 Table~\ref{tabsets}. The inversion error bars are plotted only for the
 profiles resulting from inverting Sets 1 and 2. Shaded area corresponds
 to the region where the solution of the inversion in not reliable as the
 result of the poor quality of the resolution kernels due to the lack of
 sensitivity of the acoustic modes to this region of the Sun.}
 \label{ideal}
\end{figure}

Three artificial data sets have been built (see Table~\ref{tabsets}). The
first, Set 1, is the reference and contains low-degree $p$ modes below 2.3
mHz while $p$ modes for $\ell >$ 3 extend up to 3.9 mHz. The second set, Set
2, adds the $\ell=2$ and 3 modes up to 3.4 mHz. As we have shown in the
previous section, these modes could have some sensitivity to the rotation in
the core (see Figure~\ref{sensib}a). Finally, the third set, Set 3, also
adds the $\ell=1$ modes up to 3.4 mHz that seem to be at the sensitivity limit
(see again Figure~\ref{sensib}a).

Figure~\ref{ideal} shows the resulting inversions for these three artificial data
sets. We shall qualify these inversions as {\it ideal} because all of the
splittings were given the same error bars and are noiseless. The inferred
rotation rate represents the best result that we can obtain for each set of
modes.

\begin{figure}[ht]  
   \centerline{\includegraphics[width=0.75\textwidth,clip=]{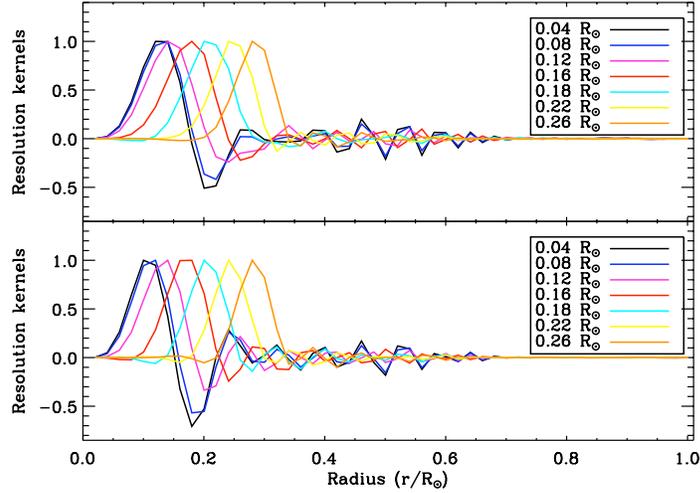} }
  \caption{Resolution kernels of the inner model grid for Set 1 (top) and 3
  (bottom) in the {\em ideal} case (all the modes have the same error bars).}
 \label{Kernideal}
\end{figure}

  As expected, the differences between the profiles appear below 0.3 $R_\odot$
but they are only significant below $\approx$ 0.15 $R_\odot$. All along the solar
interior, the differences between Set 2 and 3 are very small and are
within the inversion uncertainties. Comparing these two data sets with Set
1, we obtain an improvement up to $\approx$ 30$\%$ in the deepest region when
the sets with more modes are used. Unfortunately, in all the data sets, the
recovered profiles are not accurate enough. This is easily understood when
looking at the corresponding resolution kernels: they remain, indeed, very
broad. Figure~\ref{Kernideal} shows some resolution kernels for the deepest
model grid points and for Sets 1 and 3.  There are no resolution kernels
centered below $\sim0.13 R_\odot$ for Set 1. Comparing the resolution
kernels of Set 1 and 3, we see an improvement in the second case. Indeed
the resolution kernels are narrower (better resolution in the solution of the
inversion) and they are slightly shifted to deeper layers (by about 3$\%$ of
the solar radius).

\begin{figure}[h]  
   \centerline{\includegraphics[width=0.75\textwidth,clip=]{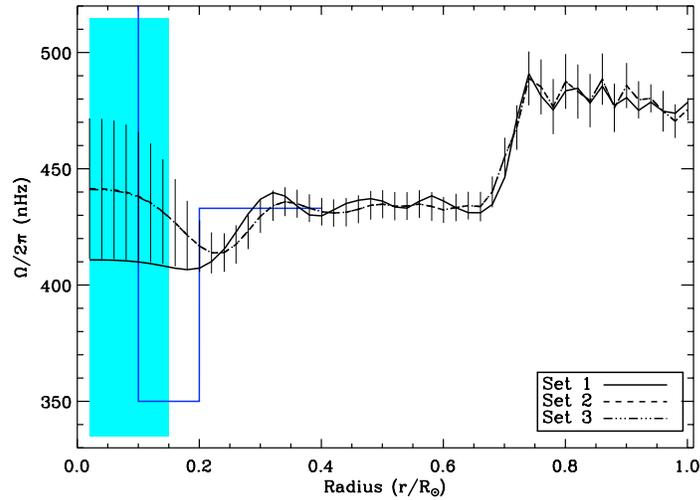} }
 \caption{Equatorial rotation rate obtained from the inversions of the
 artificial step profile (continuous line) using the three data sets
 described in Table~\ref{tabsets} but with the real error bars as described in
 Section 2. For clarity, the inversion error bars are plotted only in the
 inversion of the Set 2. Shaded area corresponds to the region where the
 solution of the inversion in not reliable as explained in
 Figure~\ref{ideal}. }
 \label{noise}
\end{figure}

We have shown the improvements of adding low-degree high-order $p$ modes in the
{\em ideal} inversions. However, the observed splittings have non-uniform
error bars. Therefore, we inverted the same sets of artificial splittings but
using the actual measured error bars. To be more realistic, Gaussian noise was
also added to the artificial splittings, commensurate to their corresponding
uncertainty.
The resulting rotation profiles are plotted in Figure~\ref{noise}. As for the
{\em ideal} case, Sets 2 and 3 give the same qualitative results and both
are slightly better in the core than Set 1, even though the differences are
within the inversion error bars. Nevertheless, the solution is improved in all
of the radiative region where the ripples are reduced and the solution is
smoother when more modes are added. Therefore, even with the realistic
uncertainties, the recovered rotation rate is more accurate when Sets 2 or
3 are used. Unfortunately, the contribution of the $\ell=1$ modes in the 2.5
to 3.4 mHz range remains negligible with the present magnitude of their
uncertainties.

\begin{figure} [htb*]
   \centerline{\includegraphics[width=0.75\textwidth,clip=]{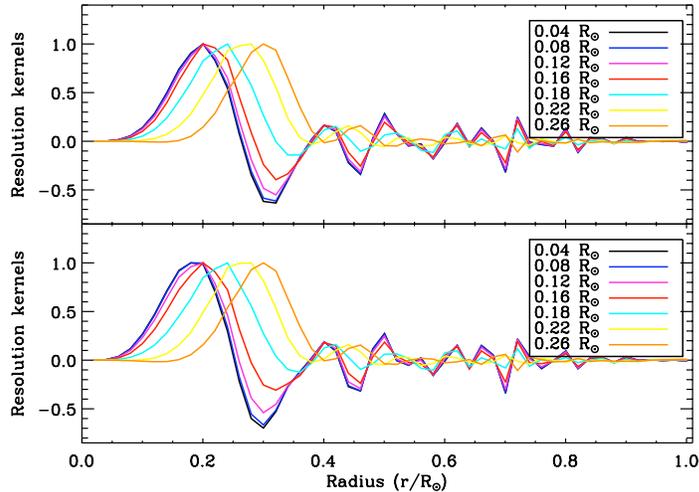} }
 \caption{Resolution kernels of the inner model grid points for Set 1 (top)
  and 3 (bottom) but using splittings with realistic uncertainties. }
  \label{Kernoise}
\end{figure}

Including realistic uncertainties results in significant changes in the
resolution kernels (see Figure~\ref{Kernoise}). Indeed, the coefficients
of the linear combination given to each splitting (inversely proportional to
the error bars) changes the linear combination of the modes used to compute
those resolution kernels. Therefore the inner resolution kernels shift
outwards: the maximum of the deepest one is now at around 0.16 $R_\odot$ for
all sets and, of course, are broader than the ones of the {\em ideal} case. The
differences between the resolution kernels deduced from Sets 1 and 3 are
smaller, explaining why the solutions in both cases are so similar.

\begin{figure}[htb*]
   \centerline{\includegraphics[width=0.75\textwidth,clip=]{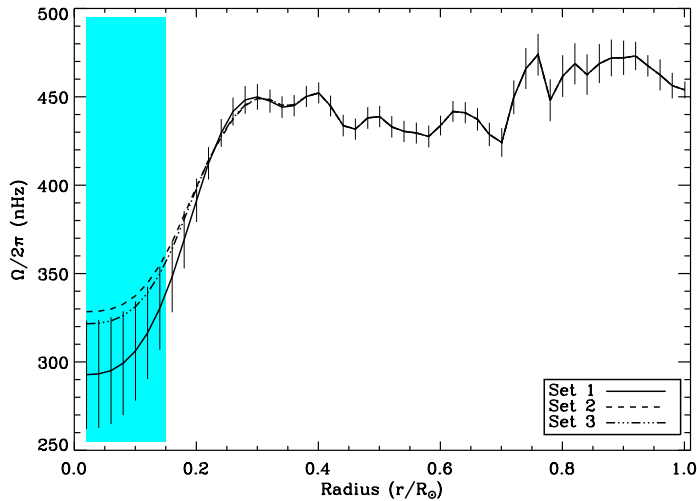} }
  \caption{Equatorial rotation rate of real data using only low-degree
  low-order $p$ modes up to 2.3 and 3.4 mHz and the same set of high-degree
  modes. For clarity, the inversion error bars are shown only for the result
  of inverting Set 1. Shaded area corresponds to the region where the solution of the
inversion in not reliable as explained in Figure~\ref{ideal}. }
  \label{real}
\end{figure}

After having studied the improvements in the rotation rate that we can expect
by adding the low-degree high-order $p$ modes, we computed the same set of
inversions using the real data. Figure~\ref{real} shows the inferred rotation
rate for the real data using the Sets 1, 2, and 3. The results are similar
to what we found for the artificial cases with realistic
uncertainties. Indeed, the rotation rate below $\approx$0.25~$R_\odot$
changes slightly when comparing Set 2 and 3 (that cannot be distinguished within
the error bars) to Set 1. The rotation profiles show a small increase inside
the core when the new modes are added (Sets 2 and 3). The difference is
barely significant down to 0.16~$R_\odot$ which is the reliability limit of
the inversion. As we already knew, we need gravity modes to get more
information about the dynamics of the deepest layers of the core
\cite{MatEff2007Inv}. In the rest of the radiative zone (above 0.25 $R_\odot$),
the profiles inferred using the three sets are the same.

We have also tested the effect of increasing the amplitude ratio of the
visible $m$ components of the $\ell=2$ and 3 modes by 20 and $10\%$
respectively. The resulting splittings are systematically increased by $\approx
10$ nHz for the $\ell=2$ modes and by 3 to 4 nHz for the $\ell=3$ modes. The
inversion computed with these new splittings gives the same qualitative
results as those shown in Figure~\ref{real}: the inferred profiles are
improved when more modes are used (Set 2 and 3). For these two profiles,
the rotation rate obtained at 0.16 $R_\odot$ has increased by some $7\%$,
which is still within the inversion error bars.

\section{Conclusion}

 We have analyzed a $4\,182$ day-long time series of GOLF velocity observation
and have derived the frequencies and the splittings of the acoustic modes up
to 3.4 mHz. We used a 45 $\mu$Hz window to fit the $\ell=0, 2$ and $1, 3$
pairs to limit any bias in the fitted splittings. By fitting shorter time
series simultaneously and including a linear dependence of the splitting with
the solar activity, we have checked that for most of the modes the variation
of this parameter with the activity is within the error bars
estimated using the whole time series.

We then carried out a study of the sensitivity of the rotation of the core
(below 0.25 $R_\odot$) to the $\ell \le 3$ modes up to 3.4 mHz. We show that
the contribution of the rotation rate in the core to the splittings of the
$\ell=2$ and 3 modes is several times the size of their current error
bars. By contrast the $\ell=1$ modes in the 2.5 to 3.4 mHz range have a
contribution from the core of the same order of magnitude as their error bars.
Therefore, they cannot provide valuable information on the core's rotation.

By using artificial inversions of a step-like profile we have shown that,
ideally, the new modes would improve the inferred rotation rate in the core
but with a small accuracy on the profile itself due to a lack of resolution in
the inner regions of the Sun -- as indicated by the corresponding resolution
kernels. The effect of including the $\ell=1$ modes between 2.5 and 3.4 mHz is
negligible with the current size of their error bars. Using the real error
bars degrades the inferred solution and the differences in the core profile
are within the inversion uncertainties. However, the profile in the rest of
the radiative region is smoother when more modes are used, demonstrating that
the inclusion of those modes improves the inversion results.

Finally, the inversion of the real data, including the larger set of
low-degree low-order $p$ modes, shows no significant differences between Set 2
and 3 as for the artificial data. There is a small increase of the
rotation rate below 0.25 $R_\odot$ compared to the reference Set 1 but
it is within the inversion error bars.  The resolution kernels are improved
when the low-degree high-order $p$ modes are included and the resolution kernels
at the deepest locations are shifted by 3$\%$ of the solar radius towards the
center. This result suggests that with the present error bars it is better to
use such modes in the inversions (at least for the high order $\ell=2$ and 3
modes).

To progress in the inference of the rotation rate in the solar core we would
need to measure low-degree modes: acoustic, mixed, and gravity modes that will
bring a direct signature of the dynamics of the still hidden core of the Sun.

\begin{acks}
The GOLF experiment is based upon a consortium of institutes (IAS, CEA/Saclay,
Nice, and Bordeaux Observatories from France, and IAC from Spain) involving a
large number of scientists and engineers, as enumerated in
\inlinecite{GabGre1995}. SOHO is a mission of international cooperation
between ESA and NASA. This work has been partially funded by the grant
AYA2004-04462 of the Spanish Ministry of Education and Culture and
partially supported by the European Helio- and Asteroseismology Network
(HELAS\footnote{\url{http://www.helas-eu.org/}}), a major international
collaboration funded by the European Commission's Sixth Framework Programme.
\end{acks}

   
\bibliographystyle{spr-mp-sola}

\bibliography{/Users/rgarcia/Desktop/BIBLIO.bib}  

\begin{thebibliography}{32}
\ifx \bisbn   \undefined \def \bisbn  #1{ISBN #1}   \fi
\ifx \binits  \undefined \def \binits#1{#1} \fi
\ifx \bauthor  \undefined \def \bauthor#1{#1} \fi
\ifx \batitle  \undefined \def \batitle#1{#1} \fi
\ifx \bjtitle  \undefined \def \bjtitle#1{#1} \fi
\ifx \bvolume  \undefined \def \bvolume#1{#1} \fi
\ifx \byear  \undefined \def \byear#1{#1} \fi
\ifx \bissue  \undefined \def \bissue#1{#1} \fi
\ifx \bfpage  \undefined \def \bfpage#1{#1} \fi
\ifx \blpage  \undefined \def \blpage #1{#1} \fi
\ifx \burl  \undefined \def \burl#1{#1} \fi
\ifx \binterref  \undefined \def \binterref#1{#1} \fi
\ifx \betal  \undefined \def \betal#1{#1} \fi
\ifx \binstitute  \undefined \def \binstitute#1{#1} \fi
\ifx \bctitle  \undefined \def \bctitle#1{#1} \fi
\ifx \beditor  \undefined \def \beditor#1{#1} \fi
\ifx \bpublisher  \undefined \def \bpublisher#1{#1} \fi
\ifx \bbtitle  \undefined \def \bbtitle#1{#1} \fi
\ifx \bedition  \undefined \def \bedition#1{#1} \fi
\ifx \bseriesno  \undefined \def \bseriesno#1{#1} \fi
\ifx \blocation  \undefined \def \blocation#1{#1} \fi
\ifx \bsertitle  \undefined \def \bsertitle#1{#1} \fi
\ifx \bsnm \undefined \def \bsnm#1{#1} \fi
\ifx \bsuffix \undefined \def \bsuffix#1{#1} \fi
\ifx \bparticle \undefined \def \bparticle#1{#1} \fi
\ifx \barticle \undefined \def \barticle#1{#1} \fi
\ifx \botherref \undefined \def \botherref #1{#1} \fi
\ifx \url \undefined \def \url#1{\textsf{#1}} \fi
\ifx \bchapter \undefined \def \bchapter#1{#1} \fi
\ifx \bbook \undefined \def \bbook#1{#1} \fi
\ifx \bcomment \undefined \def \bcomment#1{#1} \fi
\ifx \oauthor \undefined \def \oauthor#1{#1} \fi
\ifx \citeauthoryear \undefined \def \citeauthoryear#1{#1} \fi
\def \endbibitem {}

\bibitem[\protect\citeauthoryear{{Antia} and
  {Basu}}{2000}]{2000ApJ...541..442A}
\begin{barticle}
\bauthor{\bsnm{{Antia}},~\binits{H.M.}}, \bauthor{\bsnm{{Basu}},~\binits{S.}}:
\byear{2000}, \batitle{{Temporal Variations of the Rotation Rate in the Solar
  Interior}}. \textit{\bjtitle{\apj}} \textbf{\bvolume{541}},
  \bfpage{442}\,--\,\blpage{448}. \url{doi:10.1086/309421}.
\end{barticle}
\endbibitem

\bibitem[\protect\citeauthoryear{{Chaplin} {\textit{et~al.}}}{2001}]{ChaEls2001}
\begin{barticle}
\bauthor{\bsnm{{Chaplin}},~\binits{W.J.}},
  \bauthor{\bsnm{{Elsworth}},~\binits{Y.}},
  \bauthor{\bsnm{{Isaak}},~\binits{G.R.}},
  \bauthor{\bsnm{{Marchenkov}},~\binits{K.I.}},
  \bauthor{\bsnm{{Miller}},~\binits{B.A.}},
  \bauthor{\bsnm{{New}},~\binits{R.}}:
\byear{2001}, \batitle{{Rigid rotation of the solar core? On the reliable
  extraction of low-l rotational p-mode splittings from full-disc observations
  of the Sun}}. \textit{\bjtitle{\mnras}} \textbf{\bvolume{327}},
  \bfpage{1127}\,--\,\blpage{1136}.
  \binterref{\url{doi:10.1046/j.1365-8711.2001.04805.x}}.
\end{barticle}
\endbibitem

\bibitem[\protect\citeauthoryear{{Chaplin} {\textit{et~al.}}}{2004}]{ChaSek2004}
\begin{barticle}
\bauthor{\bsnm{{Chaplin}},~\binits{W.J.}},
  \bauthor{\bsnm{{Sekii}},~\binits{T.}},
  \bauthor{\bsnm{{Elsworth}},~\binits{Y.}},
  \bauthor{\bsnm{{Gough}},~\binits{D.O.}}:
\byear{2004}, \batitle{{On the detectability of a rotation-rate gradient in the
  solar core}}. \textit{\bjtitle{\mnras}} \textbf{\bvolume{355}},
  \bfpage{535}\,--\,\blpage{542}.
  \url{doi:10.1111/j.1365-2966.2004.08338.x}.
\end{barticle}
\endbibitem

\bibitem[\protect\citeauthoryear{{Chaplin} {\textit{et~al.}}}{2006}]{ChaApp2006}
\begin{barticle}
\bauthor{\bsnm{{Chaplin}},~\binits{W.J.}},
  \bauthor{\bsnm{{Appourchaux}},~\binits{T.}},
  \bauthor{\bsnm{{Baudin}},~\binits{F.}},
  \bauthor{\bsnm{{Boumier}},~\binits{P.}},
  \bauthor{\bsnm{{Elsworth}},~\binits{Y.}},
  \bauthor{\bsnm{{Fletcher}},~\binits{S.T.}},
  \bauthor{\bsnm{{Fossat}},~\binits{E.}},
  \bauthor{\bsnm{{Garc{\'{\i}}a}},~\binits{R.A.}},
  \bauthor{\bsnm{{Isaak}},~\binits{G.R.}},
  \bauthor{\bsnm{{Jim\'enez}},~\binits{A.}},
  \bauthor{\bsnm{{Jim\'enez-Reyes}},~\binits{S.J.}},
  \bauthor{\bsnm{{Lazrek}},~\binits{M.}},
  \bauthor{\bsnm{{Leibacher}},~\binits{J.W.}},
  \bauthor{\bsnm{{Lochard}},~\binits{J.}}, \bauthor{\bsnm{{New}},~\binits{R.}},
  \bauthor{\bsnm{{Pall\'e}},~\binits{P.}},
  \bauthor{\bsnm{{R\'egulo}},~\binits{C.}},
  \bauthor{\bsnm{{Salabert}},~\binits{D.}},
  \bauthor{\bsnm{{Seghouani}},~\binits{N.}},
  \bauthor{\bsnm{{Toutain}},~\binits{T.}},
  \bauthor{\bsnm{{Wachter}},~\binits{R.}}:
\byear{2006}, \batitle{{Solar FLAG hare and hounds: on the extraction of
  rotational p-mode splittings from seismic, Sun-as-a-star data}}.
  \textit{\bjtitle{\mnras}} \textbf{\bvolume{369}},
  \bfpage{985}\,--\,\blpage{996}.
  \url{doi:10.1111/j.1365-2966.2006.10358.x}.
\end{barticle}
\endbibitem

\bibitem[\protect\citeauthoryear{{Couvidat} {\textit{et~al.}}}{2003}]{CouGar2003}
\begin{barticle}
\bauthor{\bsnm{{Couvidat}},~\binits{S.}},
  \bauthor{\bsnm{{Garc{\'{\i}}a}},~\binits{R.A.}},
  \bauthor{\bsnm{{Turck-Chi\`eze}},~\binits{S.}},
  \bauthor{\bsnm{{Corbard}},~\binits{T.}},
  \bauthor{\bsnm{{Henney}},~\binits{C.J.}},
  \bauthor{\bsnm{{Jim\'enez-Reyes}},~\binits{S.}}:
\byear{2003}, \batitle{{The Rotation of the Deep Solar Layers}}.
  \textit{\bjtitle{\apjl}} \textbf{\bvolume{597}},
  \bfpage{L77}\,--\,\blpage{L79}. \url{doi:10.1086/379698}.
\end{barticle}
\endbibitem

\bibitem[\protect\citeauthoryear{{Domingo}, {Fleck}, and
  {Poland}}{1995}]{DomFle1995}
\begin{barticle}
\bauthor{\bsnm{{Domingo}},~\binits{V.}}, \bauthor{\bsnm{{Fleck}},~\binits{B.}},
  \bauthor{\bsnm{{Poland}},~\binits{A.I.}}:
\byear{1995}, \batitle{{The SOHO Mission: an Overview}}.
  \textit{\bjtitle{\solphys}} \textbf{\bvolume{162}},
  \bfpage{1}\,--\,\blpage{37}.
\end{barticle}
\endbibitem

\bibitem[\protect\citeauthoryear{{Eff-Darwich}, {Korzennik}, and
  {Jim{\'e}nez-Reyes}}{2002}]{2002ApJ...573..857E}
\begin{barticle}
\bauthor{\bsnm{{Eff-Darwich}},~\binits{A.}},
  \bauthor{\bsnm{{Korzennik}},~\binits{S.G.}},
  \bauthor{\bsnm{{Jim{\'e}nez-Reyes}},~\binits{S.J.}}:
\byear{2002}, \batitle{{Inversion of the Internal Solar Rotation Rate}}.
  \textit{\bjtitle{\apj}} \textbf{\bvolume{573}},
  \bfpage{857}\,--\,\blpage{863}. \url{doi:10.1086/340747}.
\end{barticle}
\endbibitem

\bibitem[\protect\citeauthoryear{{Eff-Darwich}
  {\textit{et~al.}}}{2007}]{effKor07}
\begin{botherref}
\oauthor{\bsnm{{Eff-Darwich}},~\binits{A.}},
  \oauthor{\bsnm{{Korzennik}},~\binits{S.G.}},
  \oauthor{\bsnm{{Jim\'enez-Reyes}},~\binits{S.J.}}, \oauthor{\bsnm{{Garc\'\i
  a}},~\binits{R.A.}}:
2007, The rotation of the solar radiative interior after 2088 days of
  helioseismic observations from GONG, GOLF and MDI. \textit{\apj}, in press.
\end{botherref}
\endbibitem

\bibitem[\protect\citeauthoryear{{Eff-Darwich} and {P\'erez
  Hern\'andez}}{1997}]{Eff-Darwich1997}
\begin{barticle}
\bauthor{\bsnm{{Eff-Darwich}},~\binits{A.}}, \bauthor{\bsnm{{P\'erez
  Hern\'andez}},~\binits{F.}}:
\byear{1997}, \batitle{A new strategy for helioseismic inversions}.
  \textit{\bjtitle{\aaps}} \textbf{\bvolume{125}}, 1.
\end{barticle}
\endbibitem

\bibitem[\protect\citeauthoryear{{Elsworth} {\textit{et~al.}}}{1995}]{ElsHow1995}
\begin{barticle}
\bauthor{\bsnm{{Elsworth}},~\binits{Y.}}, \bauthor{\bsnm{{Howe}},~\binits{R.}},
  \bauthor{\bsnm{{Isaak}},~\binits{G.R.}},
  \bauthor{\bsnm{{McLeod}},~\binits{C.P.}},
  \bauthor{\bsnm{{Miller}},~\binits{B.A.}},
  \bauthor{\bsnm{{New}},~\binits{R.}},
  \bauthor{\bsnm{{Wheeler}},~\binits{S.J.}},
  \bauthor{\bsnm{{Gough}},~\binits{D.O.}}:
\byear{1995}, \batitle{{Slow Rotation of the Sun's Interior}}.
  \textit{\bjtitle{\nat}} \textbf{\bvolume{376}}, \bfpage{669}.
  \url{doi:10.1038/376669a0}.
\end{barticle}
\endbibitem

\bibitem[\protect\citeauthoryear{{Gabriel} {\textit{et~al.}}}{1995}]{GabGre1995}
\begin{barticle}
\bauthor{\bsnm{{Gabriel}},~\binits{A.H.}},
  \bauthor{\bsnm{{Grec}},~\binits{G.}}, \bauthor{\bsnm{{Charra}},~\binits{J.}},
  \bauthor{\bsnm{{Robillot}},~\binits{J.M.}},
  \bauthor{\bsnm{{Cort\'es}},~\binits{T.R.}},
  \bauthor{\bsnm{{Turck-Chi\`eze}},~\binits{S.}},
  \bauthor{\bsnm{{Bocchia}},~\binits{R.}},
  \bauthor{\bsnm{{Boumier}},~\binits{P.}},
  \bauthor{\bsnm{{Cantin}},~\binits{M.}},
  \bauthor{\bsnm{{C\'espedes}},~\binits{E.}},
  \bauthor{\bsnm{{Cougrand}},~\binits{B.}},
  \bauthor{\bsnm{{Cretolle}},~\binits{J.}},
  \bauthor{\bsnm{{Dame}},~\binits{L.}},
  \bauthor{\bsnm{{Decaudin}},~\binits{M.}},
  \bauthor{\bsnm{{Delache}},~\binits{P.}},
  \bauthor{\bsnm{{Denis}},~\binits{N.}}, \bauthor{\bsnm{{Duc}},~\binits{R.}},
  \bauthor{\bsnm{{Dzitko}},~\binits{H.}},
  \bauthor{\bsnm{{Fossat}},~\binits{E.}},
  \bauthor{\bsnm{{Fourmond}},~\binits{J.J.}},
  \bauthor{\bsnm{{Garc{\'{\i}}a}},~\binits{R.A.}},
  \bauthor{\bsnm{{Gough}},~\binits{D.}},
  \bauthor{\bsnm{{Grivel}},~\binits{C.}},
  \bauthor{\bsnm{{Herreros}},~\binits{J.M.}},
  \bauthor{\bsnm{{Lagardere}},~\binits{H.}},
  \bauthor{\bsnm{{Moalic}},~\binits{J.P.}},
  \bauthor{\bsnm{{Pall\'e}},~\binits{P.L.}},
  \bauthor{\bsnm{{Petrou}},~\binits{N.}},
  \bauthor{\bsnm{{Sanchez}},~\binits{M.}},
  \bauthor{\bsnm{{Ulrich}},~\binits{R.}}, \bauthor{\bsnm{{van der
  Raay}},~\binits{H.B.}}:
\byear{1995}, \batitle{{Global Oscillations at Low Frequency from the SOHO
  Mission (GOLF)}}. \textit{\bjtitle{\solphys}} \textbf{\bvolume{162}},
  \bfpage{61}\,--\,\blpage{99}.
\end{barticle}
\endbibitem

\bibitem[\protect\citeauthoryear{{Garc{\'{\i}}a}
  {\textit{et~al.}}}{2004}]{GarCor2004}
\begin{barticle}
\bauthor{\bsnm{{Garc{\'{\i}}a}},~\binits{R.A.}},
  \bauthor{\bsnm{{Corbard}},~\binits{T.}},
  \bauthor{\bsnm{{Chaplin}},~\binits{W.J.}},
  \bauthor{\bsnm{{Couvidat}},~\binits{S.}},
  \bauthor{\bsnm{{Eff-Darwich}},~\binits{A.}},
  \bauthor{\bsnm{{Jim\'enez-Reyes}},~\binits{S.J.}},
  \bauthor{\bsnm{{Korzennik}},~\binits{S.G.}},
  \bauthor{\bsnm{{Ballot}},~\binits{J.}},
  \bauthor{\bsnm{{Boumier}},~\binits{P.}},
  \bauthor{\bsnm{{Fossat}},~\binits{E.}},
  \bauthor{\bsnm{{Henney}},~\binits{C.J.}},
  \bauthor{\bsnm{{Howe}},~\binits{R.}}, \bauthor{\bsnm{{Lazrek}},~\binits{M.}},
  \bauthor{\bsnm{{Lochard}},~\binits{J.}},
  \bauthor{\bsnm{{Pall{\'e}}},~\binits{P.L.}},
  \bauthor{\bsnm{{Turck-Chi\`eze}},~\binits{S.}}:
\byear{2004}, \batitle{{About the rotation of the solar radiative interior}}.
  \textit{\bjtitle{\solphys}} \textbf{\bvolume{220}},
  \bfpage{269}\,--\,\blpage{285}.
  \url{doi:10.1023/B:SOLA.0000031395.90891.ce}.
\end{barticle}
\endbibitem

\bibitem[\protect\citeauthoryear{{Garc{\'{\i}}a}
  {\textit{et~al.}}}{2004}]{2004SOHO...14..436G}
\begin{botherref}
\oauthor{\bsnm{{Garc{\'{\i}}a}},~\binits{R.A.}},
  \oauthor{\bsnm{{Jim{\'e}nez-Reyes}},~\binits{S.J.}},
  \oauthor{\bsnm{{Turck-Chi{\`e}ze}},~\binits{S.}},
  \oauthor{\bsnm{{Ballot}},~\binits{J.}},
  \oauthor{\bsnm{{Henney}},~\binits{C.J.}}:
2004, {Solar Low-Degree P-Mode Parameters after 8 Years of Velocity
  Measurements with SOHO}. In: {Danesy}, D. (ed.) \textit{ESA SP-559: SOHO 14
  Helio- and Asteroseismology: Towards a Golden Future}, \textbf{14},
  436.
\end{botherref}
\endbibitem

\bibitem[\protect\citeauthoryear{{Garc{\'{\i}}a}
  {\textit{et~al.}}}{2005}]{GarSTC2005}
\begin{barticle}
\bauthor{\bsnm{{Garc{\'{\i}}a}},~\binits{R.A.}},
  \bauthor{\bsnm{{Turck-Chi\`eze}},~\binits{S.}},
  \bauthor{\bsnm{{Boumier}},~\binits{P.}},
  \bauthor{\bsnm{{Robillot}},~\binits{J.M.}},
  \bauthor{\bsnm{{Bertello}},~\binits{L.}},
  \bauthor{\bsnm{{Charra}},~\binits{J.}},
  \bauthor{\bsnm{{Dzitko}},~\binits{H.}},
  \bauthor{\bsnm{{Gabriel}},~\binits{A.H.}},
  \bauthor{\bsnm{{Jim\'enez-Reyes}},~\binits{S.J.}},
  \bauthor{\bsnm{{Pall\'e}},~\binits{P.L.}},
  \bauthor{\bsnm{{Renaud}},~\binits{C.}}, \bauthor{\bsnm{{Roca
  Cort\'es}},~\binits{T.}}, \bauthor{\bsnm{{Ulrich}},~\binits{R.K.}}:
\byear{2005}, \batitle{{Global solar Doppler velocity determination with the
  GOLF/SoHO instrument}}. \textit{\bjtitle{\aap}} \textbf{\bvolume{442}},
  \bfpage{385}\,--\,\blpage{395}. \url{doi:10.1051/0004-6361:20052779}.
\end{barticle}
\endbibitem

\bibitem[\protect\citeauthoryear{{Garc{\'{\i}}a}
  {\textit{et~al.}}}{2007}]{2007Sci...316.1591G}
\begin{barticle}
\bauthor{\bsnm{{Garc{\'{\i}}a}},~\binits{R.A.}},
  \bauthor{\bsnm{{Turck-Chi\`eze}},~\binits{S.}},
  \bauthor{\bsnm{{Jim\'enez-Reyes}},~\binits{S.J.}},
  \bauthor{\bsnm{{Ballot}},~\binits{J.}},
  \bauthor{\bsnm{{Pall\'e}},~\binits{P.L.}},
  \bauthor{\bsnm{{Eff-Darwich}},~\binits{A.}},
  \bauthor{\bsnm{{Mathur}},~\binits{S.}},
  \bauthor{\bsnm{{Provost}},~\binits{J.}}:
\byear{2007}, \batitle{{Tracking Solar Gravity Modes: The Dynamics of the Solar
  Core}}. \textit{\bjtitle{Science}} \textbf{\bvolume{316}},
  \bfpage{1591}. \url{doi:10.1126/science.1140598}.
\end{barticle}
\endbibitem

\bibitem[\protect\citeauthoryear{{Gelly}
  {\textit{et~al.}}}{2002}]{2002A&A...394..285G}
\begin{barticle}
\bauthor{\bsnm{{Gelly}},~\binits{B.}}, \bauthor{\bsnm{{Lazrek}},~\binits{M.}},
  \bauthor{\bsnm{{Grec}},~\binits{G.}}, \bauthor{\bsnm{{Ayad}},~\binits{A.}},
  \bauthor{\bsnm{{Schmider}},~\binits{F.X.}},
  \bauthor{\bsnm{{Renaud}},~\binits{C.}},
  \bauthor{\bsnm{{Salabert}},~\binits{D.}},
  \bauthor{\bsnm{{Fossat}},~\binits{E.}}:
\byear{2002}, \batitle{{Solar p-modes from 1979 days of the GOLF experiment}}.
  \textit{\bjtitle{\aap}} \textbf{\bvolume{394}},
  \bfpage{285}\,--\,\blpage{297}. \url{doi:10.1051/0004-6361:20021106}.
\end{barticle}
\endbibitem

\bibitem[\protect\citeauthoryear{{Hansen}, {Cox}, and {van Horn}}{1977}]{bi46}
\begin{barticle}
\bauthor{\bsnm{{Hansen}},~\binits{C.J.}},
  \bauthor{\bsnm{{Cox}},~\binits{J.P.}}, \bauthor{\bsnm{{van
  Horn}},~\binits{H.M.}}:
\byear{1977}, \batitle{The effects of differential rotation on the splitting of
  nonradial modes of stellar oscillation}. \textit{\bjtitle{\apj}}
  \textbf{\bvolume{217}}, 151\,--\,159.
\end{barticle}
\endbibitem

\bibitem[\protect\citeauthoryear{{Henney}}{1999}]{1999PhDT.........8H}
\begin{botherref}
\oauthor{\bsnm{{Henney}},~\binits{C.J.}}:
1999, Comparison between simultaneous golf and mdi observations in search of
  low frequency solar oscillations.
PhD thesis, University of California, Los Angeles.
\end{botherref}
\endbibitem

\bibitem[\protect\citeauthoryear{{Howe} {\textit{et~al.}}}{2000}]{HowJCD2000}
\begin{barticle}
\bauthor{\bsnm{{Howe}},~\binits{R.}},
  \bauthor{\bsnm{{Christensen-Dalsgaard}},~\binits{J.}},
  \bauthor{\bsnm{{Hill}},~\binits{F.}}, \bauthor{\bsnm{{Komm}},~\binits{R.W.}},
  \bauthor{\bsnm{{Larsen}},~\binits{R.M.}},
  \bauthor{\bsnm{{Schou}},~\binits{J.}},
  \bauthor{\bsnm{{Thompson}},~\binits{M.J.}},
  \bauthor{\bsnm{{Toomre}},~\binits{J.}}:
\byear{2000}, \batitle{{Dynamic Variations at the Base of the Solar Convection
  Zone}}. \textit{\bjtitle{Science}} \textbf{\bvolume{287}},
  \bfpage{2456}\,--\,\blpage{2460}.
\end{barticle}
\endbibitem

\bibitem[\protect\citeauthoryear{{Jim\'enez}
  {\textit{et~al.}}}{1994}]{1994ApJ...435..874J}
\begin{barticle}
\bauthor{\bsnm{{Jim\'enez}},~\binits{A.}}, \bauthor{\bsnm{{P\'erez
  Hernandez}},~\binits{F.}}, \bauthor{\bsnm{{Claret}},~\binits{A.}},
  \bauthor{\bsnm{{Pall\'e}},~\binits{P.L.}},
  \bauthor{\bsnm{{R\'egulo}},~\binits{C.}}, \bauthor{\bsnm{{Roca
  Cort\'es}},~\binits{T.}}:
\byear{1994}, \batitle{{The rotation of the solar core}}.
  \textit{\bjtitle{\apj}} \textbf{\bvolume{435}},
  \bfpage{874}\,--\,\blpage{880}. \url{doi:10.1086/174868}.
\end{barticle}
\endbibitem

\bibitem[\protect\citeauthoryear{{Jim{\'e}nez-Reyes}
  {\textit{et~al.}}}{2001}]{2001A&A...379..622J}
\begin{barticle}
\bauthor{\bsnm{{Jim{\'e}nez-Reyes}},~\binits{S.J.}},
  \bauthor{\bsnm{{Corbard}},~\binits{T.}},
  \bauthor{\bsnm{{Pall{\'e}}},~\binits{P.L.}}, \bauthor{\bsnm{{Roca
  Cort{\'e}s}},~\binits{T.}}, \bauthor{\bsnm{{Tomczyk}},~\binits{S.}}:
\byear{2001}, \batitle{{Analysis of the solar cycle and core rotation using 15
  years of Mark-I observations: 1984-1999. I. The solar cycle}}.
  \textit{\bjtitle{\aap}} \textbf{\bvolume{379}},
  \bfpage{622}\,--\,\blpage{633}. \url{doi:10.1051/0004-6361:20011374}.
\end{barticle}
\endbibitem

\bibitem[\protect\citeauthoryear{{Jim{\'e}nez}, {Roca Cort{\'e}s}, and
  {Jim{\'e}nez-Reyes}}{2002}]{2002SoPh..209..247J}
\begin{barticle}
\bauthor{\bsnm{{Jim{\'e}nez}},~\binits{A.}}, \bauthor{\bsnm{{Roca
  Cort{\'e}s}},~\binits{T.}},
  \bauthor{\bsnm{{Jim{\'e}nez-Reyes}},~\binits{S.J.}}:
\byear{2002}, \batitle{{Variation of the low-degree solar acoustic mode
  parameters over the solar cycle}}. \textit{\bjtitle{\solphys}}
  \textbf{\bvolume{209}}, \bfpage{247}\,--\,\blpage{263}.
\end{barticle}
\endbibitem

\bibitem[\protect\citeauthoryear{{Korzennik}}{2005}]{Kor2005}
\begin{barticle}
\bauthor{\bsnm{{Korzennik}},~\binits{S.G.}}:
\byear{2005}, \batitle{{A Mode-Fitting Methodology Optimized for Very Long
  Helioseismic Time Series}}. \textit{\bjtitle{\apj}} \textbf{\bvolume{626}},
  \bfpage{585}\,--\,\blpage{615}. \url{doi:10.1086/429748}.
\end{barticle}
\endbibitem

\bibitem[\protect\citeauthoryear{{Lopes} and
  {Turck-Chi\`eze}}{1994}]{1994A&A...290..845L}
\begin{barticle}
\bauthor{\bsnm{{Lopes}},~\binits{I.}},
  \bauthor{\bsnm{{Turck-Chi\`eze}},~\binits{S.}}:
\byear{1994}, \batitle{{The second order asymptotic theory for the solar and
  stellar low degree acoustic mode predictions}}. \textit{\bjtitle{\aap}}
  \textbf{\bvolume{290}}, \bfpage{845}\,--\,\blpage{860}.
\end{barticle}
\endbibitem

\bibitem[\protect\citeauthoryear{{Mathur}
  {\textit{et~al.}}}{2007}]{MatEff2007Inv}
\begin{botherref}
\oauthor{\bsnm{{Mathur}},~\binits{S.}},
  \oauthor{\bsnm{{Eff-Darwich}},~\binits{A.M.}},
  \oauthor{\bsnm{{Garc{\'{\i}}a}},~\binits{R.A.}},
  \oauthor{\bsnm{{Turck-Chi\`eze}},~\binits{S.}}:
2007, Sensitivity of helioseismic gravity modes to the dynamics of the solar
  core. \textit{\aap}, in press.
\end{botherref}
\endbibitem

\bibitem[\protect\citeauthoryear{{Nigam} and
  {Kosovichev}}{1998}]{1998ApJ...505L..51N}
\begin{barticle}
\bauthor{\bsnm{{Nigam}},~\binits{R.}},
  \bauthor{\bsnm{{Kosovichev}},~\binits{A.G.}}:
\byear{1998}, \batitle{{Measuring the Sun's Eigenfrequencies from Velocity and
  Intensity Helioseismic Spectra: Asymmetrical Line Profile-fitting Formula}}.
  \textit{\bjtitle{\apjl}} \textbf{\bvolume{505}}, L51.
  \url{doi:10.1086/311594}.
\end{barticle}
\endbibitem

\bibitem[\protect\citeauthoryear{{Provost}, {Berthomieu}, and
  {Morel}}{2000}]{ProBer2000}
\begin{barticle}
\bauthor{\bsnm{{Provost}},~\binits{J.}},
  \bauthor{\bsnm{{Berthomieu}},~\binits{G.}},
  \bauthor{\bsnm{{Morel}},~\binits{P.}}:
\byear{2000}, \batitle{{Low-frequency p- and g-mode solar oscillations}}.
  \textit{\bjtitle{\aap}} \textbf{\bvolume{353}},
  \bfpage{775}\,--\,\blpage{785}.
\end{barticle}
\endbibitem

\bibitem[\protect\citeauthoryear{{Roca Cort{\'e}s}
  {\textit{et~al.}}}{1998}]{1998ESASP.418..323R}
\begin{botherref}
\oauthor{\bsnm{{Roca Cort{\'e}s}},~\binits{T.}},
  \oauthor{\bsnm{{Lazrek}},~\binits{M.}},
  \oauthor{\bsnm{{Bertello}},~\binits{L.}},
  \oauthor{\bsnm{{Thiery}},~\binits{S.}},
  \oauthor{\bsnm{{Baudin}},~\binits{F.}},
  \oauthor{\bsnm{{Boumier}},~\binits{P.}},
  \oauthor{\bsnm{{Gavryusev}},~\binits{V.}},
  \oauthor{\bsnm{{Garcia}},~\binits{R.A.}},
  \oauthor{\bsnm{{Regulo}},~\binits{C.}},
  \oauthor{\bsnm{{Ulrich}},~\binits{R.K.}},
  \oauthor{\bsnm{{Grec}},~\binits{G.}}, \oauthor{\bsnm{{the GOLF Team}},~}:
1998, {The Solar Acoustic Spectrum as Seen by GOLF. II. Noise Statistics
  Background and Methods of Analysis}. In: {Korzennik}, S. (ed.)
  \textit{Structure and Dynamics of the Interior of the Sun and Sun-like
  Stars}, \textit{ESA Special Publication}, \textbf{418}, 323.
\end{botherref}
\endbibitem

\bibitem[\protect\citeauthoryear{{Scherrer}
  {\textit{et~al.}}}{1995}]{1995SoPh..162..129S}
\begin{barticle}
\bauthor{\bsnm{{Scherrer}},~\binits{P.H.}},
  \bauthor{\bsnm{{Bogart}},~\binits{R.S.}},
  \bauthor{\bsnm{{Bush}},~\binits{R.I.}},
  \bauthor{\bsnm{{Hoeksema}},~\binits{J.T.}},
  \bauthor{\bsnm{{Kosovichev}},~\binits{A.G.}},
  \bauthor{\bsnm{{Schou}},~\binits{J.}},
  \bauthor{\bsnm{{Rosenberg}},~\binits{W.}},
  \bauthor{\bsnm{{Springer}},~\binits{L.}},
  \bauthor{\bsnm{{Tarbell}},~\binits{T.D.}},
  \bauthor{\bsnm{{Title}},~\binits{A.}},
  \bauthor{\bsnm{{Wolfson}},~\binits{C.J.}},
  \bauthor{\bsnm{{Zayer}},~\binits{I.}}, \bauthor{\bsnm{{MDI Engineering
  Team}},~}:
\byear{1995}, \batitle{{The Solar Oscillations Investigation - Michelson
  Doppler Imager}}. \textit{\bjtitle{\solphys}} \textbf{\bvolume{162}},
  \bfpage{129}\,--\,\blpage{188}. \url{doi:10.1007/BF00733429}.
\end{barticle}
\endbibitem

\bibitem[\protect\citeauthoryear{{Schou}
  {\textit{et~al.}}}{1998}]{1998ApJ...505..390S}
\begin{barticle}
\bauthor{\bsnm{{Schou}},~\binits{J.}}, \bauthor{\bsnm{{Antia}},~\binits{H.M.}},
  \bauthor{\bsnm{{Basu}},~\binits{S.}},
  \bauthor{\bsnm{{Bogart}},~\binits{R.S.}},
  \bauthor{\bsnm{{Bush}},~\binits{R.I.}},
  \bauthor{\bsnm{{Chitre}},~\binits{S.M.}},
  \bauthor{\bsnm{{Christensen-Dalsgaard}},~\binits{J.}}, \bauthor{\bsnm{{di
  Mauro}},~\binits{M.P.}}, \bauthor{\bsnm{{Dziembowski}},~\binits{W.A.}},
  \bauthor{\bsnm{{Eff-Darwich}},~\binits{A.}},
  \bauthor{\bsnm{{Gough}},~\binits{D.O.}},
  \bauthor{\bsnm{{Haber}},~\binits{D.A.}},
  \bauthor{\bsnm{{Hoeksema}},~\binits{J.T.}},
  \bauthor{\bsnm{{Howe}},~\binits{R.}},
  \bauthor{\bsnm{{Korzennik}},~\binits{S.G.}},
  \bauthor{\bsnm{{Kosovichev}},~\binits{A.G.}},
  \bauthor{\bsnm{{Larsen}},~\binits{R.M.}},
  \bauthor{\bsnm{{Pijpers}},~\binits{F.P.}},
  \bauthor{\bsnm{{Scherrer}},~\binits{P.H.}},
  \bauthor{\bsnm{{Sekii}},~\binits{T.}},
  \bauthor{\bsnm{{Tarbell}},~\binits{T.D.}},
  \bauthor{\bsnm{{Title}},~\binits{A.M.}},
  \bauthor{\bsnm{{Thompson}},~\binits{M.J.}},
  \bauthor{\bsnm{{Toomre}},~\binits{J.}}:
\byear{1998}, \batitle{{Helioseismic Studies of Differential Rotation in the
  Solar Envelope by the Solar Oscillations Investigation Using the Michelson
  Doppler Imager}}. \textit{\bjtitle{\apj}} \textbf{\bvolume{505}},
  \bfpage{390}\,--\,\blpage{417}. \url{doi:10.1086/306146}.
\end{barticle}
\endbibitem

\bibitem[\protect\citeauthoryear{{Thompson} {\textit{et~al.}}}{1996}]{ThoTOO1996}
\begin{barticle}
\bauthor{\bsnm{{Thompson}},~\binits{M.J.}},
  \bauthor{\bsnm{{Toomre}},~\binits{J.}},
  \bauthor{\bsnm{{Anderson}},~\binits{E.}},
  \bauthor{\bsnm{{Antia}},~\binits{H.M.}},
  \bauthor{\bsnm{{Berthomieu}},~\binits{G.}},
  \bauthor{\bsnm{{Burtonclay}},~\binits{D.}},
  \bauthor{\bsnm{{Chitre}},~\binits{S.M.}},
  \bauthor{\bsnm{{Christensen-Dalsgaard}},~\binits{J.}},
  \bauthor{\bsnm{{Corbard}},~\binits{T.}},
  \bauthor{\bsnm{{Derosa}},~\binits{M.}},
  \bauthor{\bsnm{{Genovese}},~\binits{C.R.}},
  \bauthor{\bsnm{{Gough}},~\binits{D.O.}},
  \bauthor{\bsnm{{Haber}},~\binits{D.A.}},
  \bauthor{\bsnm{{Harvey}},~\binits{J.W.}},
  \bauthor{\bsnm{{Hill}},~\binits{F.}}, \bauthor{\bsnm{{Howe}},~\binits{R.}},
  \bauthor{\bsnm{{Korzennik}},~\binits{S.G.}},
  \bauthor{\bsnm{{Kosovichev}},~\binits{A.G.}},
  \bauthor{\bsnm{{Leibacher}},~\binits{J.W.}},
  \bauthor{\bsnm{{Pijpers}},~\binits{F.P.}},
  \bauthor{\bsnm{{Provost}},~\binits{J.}},
  \bauthor{\bsnm{{Rhodes}},~\binits{E.J.}},
  \bauthor{\bsnm{{Schou}},~\binits{J.}}, \bauthor{\bsnm{{Sekii}},~\binits{T.}},
  \bauthor{\bsnm{{Stark}},~\binits{P.B.}},
  \bauthor{\bsnm{{Wilson}},~\binits{P.}}:
\byear{1996}, \batitle{{Differential Rotation and Dynamics of the Solar
  Interior}}. \textit{\bjtitle{Science}} \textbf{\bvolume{272}},
  \bfpage{1300}\,--\,\blpage{1305}.
\end{barticle}
\endbibitem

\bibitem[\protect\citeauthoryear{{Toutain} and
  {Appourchaux}}{1994}]{1994A&A...289..649T}
\begin{barticle}
\bauthor{\bsnm{{Toutain}},~\binits{T.}},
  \bauthor{\bsnm{{Appourchaux}},~\binits{T.}}:
\byear{1994}, \batitle{{Maximum likelihood estimators: AM application to the
  estimation of the precision of helioseismic measurements}}.
  \textit{\bjtitle{\aap}} \textbf{\bvolume{289}},
  \bfpage{649}\,--\,\blpage{658}.
\end{barticle}
\endbibitem

\end{thebibliography}

\IfFileExists{\jobname.bbl}{} {\typeout{}
\typeout{****************************************************}
\typeout{****************************************************}
\typeout{** Please run "bibtex \jobname" to obtain} \typeout{**
the bibliography and then re-run LaTeX} \typeout{** twice to fix
the references !}
\typeout{****************************************************}
\typeout{****************************************************}
\typeout{}}

\end{article} 
\end{document}